\documentclass[11pt,a4paper]{article}
\pdfoutput=1
\usepackage{jheppub}
\usepackage{tikz}
\usetikzlibrary{intersections, calc, arrows.meta}
\usepackage{subcaption}
\usepackage{bm}
\usepackage{bbold}

\DeclareMathOperator{\Tr}{Tr}

\newcommand{\ri}{\mathrm{i}}
\renewcommand{\th}{\theta}

\newcommand{\cob}{\delta}
\newcommand{\ep}{\epsilon}

\newcommand{\hf}{\frac{1}{2}}

\newcommand{\del}{\partial}

\newcommand{\bra}{\langle}
\newcommand{\ket}{\rangle}
\newcommand{\la}{\lambda}

\newcommand{\bt}{\beta}
\newcommand{\ga}{\gamma}

\newcommand{\al}{\alpha}

\newcommand{\cO}{\mathcal{O}}

\newcommand{\cH}{\mathcal{H}}

\newcommand{\cN}{\mathcal{N}}

\newcommand{\cE}{\mathcal{E}}

\makeatletter
\gdef\@fpheader{}
\makeatother
\begin{document}

\title{Negativity and its capacity in JT gravity}

\author{Kazumi Okuyama and Takeshi Tachibana}

\affiliation{Department of Physics, 
Shinshu University, 3-1-1 Asahi, Matsumoto 390-8621, Japan}

\emailAdd{kazumi@azusa.shinshu-u.ac.jp, 22HS305D@shinshu-u.ac.jp}

\abstract{
We study the refined R\'{e}nyi negativity in the matrix model of Jackiw-Teitelboim (JT) gravity.
We first consider the JT gravity with dynamical branes,
which serves as a toy model of the evaporating black hole.
By including the backreaction of branes, 
we find that the refined R\'{e}nyi negativity 
monotonically decreases at late time of the evaporation.
Next we define a novel quantity, which we call ``capacity of negativity,'' as a 
derivative of the refined R\'{e}nyi negativity with respect to the replica number.
We find that the capacity of negativity has two peaks as a function of time, 
which comes from the exchange of dominance of 
the different types of replica wormholes.
}

\maketitle

\section{Introduction}
Recent developments in the study of the black hole information paradox \cite{Hawking:1976ra}
suggest a nontrivial connection between the interior and the exterior of black holes.
This connection is realized as the island in the quantum extremal surface 
formula
\cite{Penington:2019npb, Almheiri:2019psf, Almheiri:2019hni}
which is essential in reproducing the unitary Page curve \cite{Page:1993df, Page:1993wv}
of the entanglement entropy of the Hawking radiation.
In the gravitational computation of the entanglement entropy,
the replica wormholes \cite{Almheiri:2019qdq, Penington:2019kki} play a crucial role.
See \cite{Almheiri:2020cfm} for a recent review of these developments and references therein.

As another development,
the matrix model of Jackiw-Teitelboim (JT) gravity \cite{Jackiw:1984je, Teitelboim:1983ux}
was discovered by Saad, Shenker and Stanford \cite{Saad:2019lba}
and it opened a new avenue to the detailed study of the quantum gravity in lower dimensions.
JT gravity matrix model is a special case of the two dimensional topological gravity \cite{Witten:1990hr, Kontsevich:1992ti, Dijkgraaf:2018vnm}
obtained by setting the coupling of topological gravity to a specific value \cite{Okuyama:2019xbv}.
In \cite{Penington:2019kki} the Page curve is reproduced from the replica computation
in JT gravity with the end of the world (EOW) branes,
which represents the interior partners of the Hawking radiation.
In \cite{Penington:2019kki} the branes were treated as non-dynamical objects.
However,
we cannot neglect the backreaction of branes at late time of the black hole evaporation.
In \cite{Okuyama:2021bqg} the Page curve was studied by treating the branes as dynamical objects.
More specifically, the authors of \cite{Okuyama:2021bqg} studied the matrix model of JT gravity
with dynamical Fateev-Zamolodchikov-Zamolodchikov-Teschner (FZZT) \cite{Fateev:2000ik, Teschner:2000md} antibranes
as a model of the evaporating black hole.
It was found that the entanglement entropy
decreases monotonically at late time if we take account of the backreaction of branes \cite{Okuyama:2021bqg},
while it approaches a constant in the original non-dynamical treatment of branes \cite{Penington:2019kki}.

In this paper 
we consider the entanglement negativity \cite{Vidal:2002zz}
which is a measure of entanglement in general mixed states.
As shown in \cite{Dong:2021oad}, the entanglement negativity and its R\'{e}nyi generalizations 
exhibit more elaborate phase structure of the replica wormholes compared to the entanglement entropy.
Interestingly, 
there is a phase where the dominant replica wormhole is not invariant under the permutation of the replicas 
and the replica symmetry is spontaneously broken.
Note that in \cite{Dong:2021oad}
the entanglement negativity was studied in the model of JT gravity with the EOW branes \cite{Penington:2019kki}
and the topological model with the EOW branes \cite{Marolf:2020xie}, 
where the branes were treated as non-dynamical objects.
One of the purpose of this paper is to study the effect of the backreaction 
of branes to the entanglement negativity 
and its R\'{e}nyi generalization, known as the refined R\'{e}nyi negativity.
We find a monotonically decreasing behavior of the refined R\'{e}nyi 
negativity at late time of the black hole evaporation.
It turns out that this decreasing behavior can be easily understood
by using the relation between the refined R\'{e}nyi negativity
and the refined R\'{e}nyi entropy \cite{Dong:2016fnf}, as we will explain in section \ref{Sec:Page}.
 
Also, in section \ref{Sec:capacity} we study a novel quantity which we call the ``capacity of negativity.''
The capacity of negativity is a natural analogue of the capacity of entanglement \cite{Yao:2010woi},
and it is defined by the derivative of the refined R\'{e}nyi negativity with respect to the replica number.
As shown in \cite{Kawabata:2021hac, Okuyama:2021ylc}, the capacity of entanglement has a peak around the Page time 
which indicates that there is a phase transition from the disconnected saddle to the totally connected one.
In the case of the capacity of negativity,
we find two peaks around the phase transitions,
reflecting the fact that the entanglement negativity has more elaborate phase structure than that of the entanglement entropy.
Moreover, 
we find that, away from the two peaks,
the capacity of negativity approaches a universal constant value which only depends on the replica number.

This paper is organized as follows.
In section \ref{Sec:negativity}, we briefly review the entanglement negativity and its R\'{e}nyi generalizations.
We introduce the capacity of negativity as a derivative of the refined R\'{e}nyi negativity.
In section \ref{Sec:JT}, we describe our formalism to compute the effect of the backreaction to the entanglement negativity.
The deformation of the density of states due to branes plays an essential role in this computation.
In section \ref{Sec:Page}, we study the refined R\'{e}nyi negativity numerically.
In a certain limit, we analytically prove the monotonically decreasing behavior at late time of the evaporation
by using the relation between the refined R\'{e}nyi negativity and the refined R\'{e}nyi entropy.
In section \ref{Sec:capacity}, we study the capacity of negativity numerically in the microcanonical ensemble.
In the microcanonical ensemble, Schwinger-Dyson equation for the resolvent can be solved analytically
and it allows us to compute the capacity of negativity numerically.
Finally we conclude in section \ref{Sec:conclusion}.

\section{Basics of the entanglement negativity}\label{Sec:negativity}
In this section, we briefly review the entanglement negativity and its R\'{e}nyi generalizations.
We also define a novel quantity which we call the 
``capacity of negativity'' as an analogue of the capacity of entanglement.

\subsection{Partial transpose}
The entanglement negativity \cite{Vidal:2002zz} is a measure of entanglement in general mixed states,
which is computed from a partially transposed density matrix \cite{Peres:1996dw, Horodecki:1996nc}  of a bipartite quantum system
with the Hilbert space $\cH = \cH_\mathrm{A} \otimes \cH_\mathrm{B}$.
For a density matrix $\rho_\mathrm{AB}$ of the total system $\mathrm{AB}$,
we define the partially transposed density matrix $\rho_\mathrm{AB}^{\mathrm{T_B}}$ by
\begin{equation}
    \bra a, b | \rho_\mathrm{AB}^{\mathrm{T_B}} | a', b' \ket = \bra a, b' | \rho_\mathrm{AB} | a', b \ket,
\end{equation}
where $\mathrm{T_B}$ denotes the transpose on the subsystem B.
Recall that the state $\rho_\mathrm{AB}$ is called a separable state if $\rho_\mathrm{AB}$ can be factorized as
\begin{equation}
\rho_\mathrm{AB} = \sum_m p_m \rho_\mathrm{A}^{(m)} \otimes \rho_\mathrm{B}^{(m)}, ~~~ \sum_m p_m = 1,\quad p_m\geq0.
\label{eq:separable}
\end{equation}
In this case, all eigenvalues of $\rho^{\mathrm{T_B}}_{\mathrm{AB}}
=\sum_m p_m \rho_\mathrm{A}^{(m)} \otimes (\rho_\mathrm{B}^{(m)})^\mathrm{T}$ are non-negative
since every $(\rho^{(m)}_{\mathrm{B}})^\mathrm{T}$ is non-negative by definition of the density matrix
$\rho^{(m)}_{\mathrm{B}}$.\footnote{
As proven in \cite{Horodecki:1996nc}, for $2 \times 2$ and $2 \times 3$ matrices,
the non-negativity of $\rho_{AB}^{\mathrm{T_B}}$ is both necessary and sufficient condition for the state to be separable.
However it is not sufficient for more larger matrices.}
However, for an entangled, non-separable state some of the eigenvalues of $\rho^{\mathrm{T_B}}_{\mathrm{AB}}$ can be negative.
This suggests that one can define useful measures of entanglement from the eigenvalues of $\rho^{\mathrm{T_B}}_{\mathrm{AB}}$. 
For instance, the entanglement negativity is defined by
\begin{equation}
    \cN =  \sum_i \frac{|\la_i| - \la_i}{2} = \sum_{\la_i <0} |\la_i|,
\label{eq:def-N}
\end{equation}
where $\la_i$'s are the eigenvalues of $\rho_\mathrm{AB}^{\mathrm{T_B}}$.
Similarly, the logarithmic negativity is defined by
\begin{equation}
    \cE = \log \left( \sum_i |\la_i| \right) = \log \left( 2 \cN + 1 \right).
\label{eq:def-E}
\end{equation}
In the last equality of \eqref{eq:def-E}, we used the relation
\begin{equation}
 \sum_i\la_i=\Tr \rho^{\mathrm{T_B}}_{\mathrm{AB}}=\Tr \rho_{\mathrm{AB}}=1.
\label{eq:sum-la}
\end{equation} 
It is known that the logarithmic negativity sets the upper bound of the distillable entanglement
\cite{Vidal:2002zz, Plenio:2005cwa, audenaert2002entanglement}.
Note that $\cN$ and $\cE$ vanish if all $\la_i$'s are non-negative.
In particular, for a separable state $\cN=\cE=0$.
This implies that when $\cN$ and $\cE$ are non-zero the system is not in a separable state. 
Thus we can definitely say that the system A and the system B are entangled when $\cN$ and $\cE$ are non-zero.

\subsection{R\'{e}nyi generalization}
Now let us consider the R\'{e}nyi generalization of the entanglement negativity.
The $n$-th R\'{e}nyi negativity
\footnote{In \cite{Dong:2021oad}, the R\'{e}nyi negativity is denoted by $\cN_n$.
In this paper, we denote it by $Z^{\mathrm{T_B}(n)}$ so that the analogy in Table \ref{analogy} is manifest.} is defined by
\begin{equation}\label{Renyi}
    Z^{\mathrm{T_B}(n)} = \Tr \left[ \left( \rho_\mathrm{AB}^{\mathrm{T_B}} \right)^n \right] = \sum_i (\la_i)^n.
\end{equation}
Since the entanglement negativity $\cN$ in \eqref{eq:def-N}
and the logarithmic negativity $\cE$ 
in \eqref{eq:def-E}
are defined by the absolute values of the eigenvalues,
we take different analytic continuations for even and odd replica number $n$
\begin{align}
    Z^{\mathrm{T_B}(2m, \text{even})} &= \sum_{i} | \la_i |^{2m}, \label{even-Renyi} \\
    Z^{\mathrm{T_B}(2m-1, \text{odd})} &= \sum_{i} \mathrm{sgn} (\la_i)  | \la_i |^{2m-1} \label{odd-Renyi}.
\end{align}
Then $\cE$ in \eqref{eq:def-E} is obtained by the following limit
\begin{equation}
    \cE = \lim_{m \rightarrow \hf} \log Z^{\mathrm{T_B}(2m, \text{even})}.
\end{equation}
In this paper, we focus on the refined R\'{e}nyi negativities \cite{Dong:2021oad}
\footnote{
Note that $S^{\mathrm{T_B}} = \lim_{m \rightarrow 1} S^{\mathrm{T_B} (2m-1, \text{odd})}$
is called the ``odd entanglement entropy'' or the ``partially transposed entropy'' \cite{Tamaoka:2018ned, Dong:2021clv}
since its definition is the same as the von Neumann entropy with $\rho_{\mathrm{AB}}$
replaced by $\rho_{\mathrm{AB}}^{\mathrm{T_B}}$.
}
\begin{align}
    S^{\mathrm{T_B} (n, \text{even})}
    &= - n^2 \del_n \left( \frac{1}{n} \log Z^{\mathrm{T_B}(n, \text{even})} \right), \label{even-pS} \\
    S^{\mathrm{T_B} (n, \text{odd})}
    &= - n^2 \del_n \left( \frac{1}{n} \log Z^{\mathrm{T_B}(n, \text{odd})} \right), \label{odd-pS}
\end{align}
which are defined in analogy with the refined R\'{e}nyi entropy \cite{Dong:2016fnf}.
In section \ref{Sec:Page}, we consider the refined R\'{e}nyi negativities in JT gravity
and study the effect of the backreaction of branes to these quantities.

Note that there is a natural correspondence
among the statistical mechanical quantities, the R\'{e}nyi entropic quantities,
and the R\'{e}nyi negativities under the appropriate identifications \cite{Qi,Dong:2016fnf, Kawabata:2021vyo} (see Table \ref{analogy}).
\begin{table}[h]
    \renewcommand{\arraystretch}{1.5}
	\begin{center}
	\begin{tabular}{cccc}
    \hline
	\textbf{Statistical mechanics}  &\textbf{R{\'e}nyi entropic quantities}  &\textbf{R{\'e}nyi negativities} \\ \hline \hline
{\footnotesize \begin{tabular}{l}
    $\displaystyle \bt$ \\
    $H$ \\
    $\displaystyle Z(\beta) = \Tr\left[ e^{-\beta\,H}\right]$ \\ 
    $\displaystyle F(\beta) = - \beta^{-1}\log Z(\beta)$\\
    $\displaystyle E(\beta) = -\partial_\beta \log Z(\beta)$\\
    $\displaystyle S(\beta) = \beta^2\, \partial_\beta F(\beta)$\\
    $\displaystyle C(\beta) = - \beta\, \partial_\beta S(\beta)$
\end{tabular}}
&{\footnotesize \begin{tabular}{l}
    $\displaystyle n$\\
    $\displaystyle H_\mathrm{A}= -\log\rho_\mathrm{A}$\\
    $\displaystyle Z(n) = \Tr_\mathrm{A}\left[ e^{-n\,H_\mathrm{A}}\right]$ \\
    $\displaystyle F(n) = - n^{-1}\log Z(n)$\\
    $\displaystyle E(n) = -\partial_n \log Z(n)$\\
    $\displaystyle S(n) = n^2\, \partial_n F(n)$\\
    $\displaystyle C(n) = - n\, \partial_n S(n)$
\end{tabular}}
&{\footnotesize \begin{tabular}{l}
    $\displaystyle n$\\
    $\displaystyle H^\mathrm{T_B}= -\log\rho_{AB}^\mathrm{T_B}$\\
    $\displaystyle Z^{\mathrm{T_B}(n)} = \Tr\left[ e^{-n\,H^\mathrm{T_B}}\right]$ \\
    $\displaystyle F^{\mathrm{T_B}(n)} = - n^{-1}\log Z^{\mathrm{T_B}(n)}$\\
    $\displaystyle E^{\mathrm{T_B}(n)} = -\partial_n \log Z^{\mathrm{T_B}(n)}$\\
    \color{blue}{$\displaystyle S^{\mathrm{T_B}(n)} = n^2\, \partial_n F^{\mathrm{T_B}(n)}$}\\
    \color{blue}{$\displaystyle C^{\mathrm{T_B}(n)} = - n\, \partial_n S^{\mathrm{T_B}(n)}$}
\end{tabular}} \\
    \hline
	\end{tabular}
	\end{center}
\caption{
The correspondence among the statistical mechanical quantities, the R\'{e}nyi entropic quantities, and the R\'{e}nyi negativities.
Here $\bt$ is the inverse temperature, $n$ is the replica number,
$\rho_\mathrm{A}$ is the reduced density matrix and $H_\mathrm{A}$ is the modular Hamiltonian.
In the third column,
$H^\mathrm{T_B}$ denotes the partially transposed version of the modular Hamiltonian and 
we ignored the difference between even $n$ and odd $n$ for simplicity.}
\label{analogy}
\end{table}
By using the correspondence in Table \ref{analogy},
we define novel quantities which we call the ``capacity of negativities'' 
\begin{align}
    C^{\mathrm{T_B} (n, \text{even})} &= - n \del_n S^{\mathrm{T_B} (n, \text{even})}, \label{even-pC} \\
    C^{\mathrm{T_B} (n, \text{odd})} &= - n \del_n S^{\mathrm{T_B} (n, \text{odd})}. \label{odd-pC}
\end{align}
As shown in \cite{Kawabata:2021hac, Okuyama:2021ylc}, the capacity of entanglement has a peak around the Page time,
which comes from the phase transition between different topologies of the replica wormholes.
At the Page time, the disconnected saddle and the totally connected one exchange dominance,
which is the physical origin of the Page curve of the entanglement entropy.
In the case of the entanglement negativity and its R\'{e}nyi generalizations,
there appears an additional type of saddle, the partially connected one.
As emphasized in \cite{Dong:2021oad}, in the phase where 
the partially connected saddle is dominant the replica symmetry is spontaneously broken.
From the existence of the partially connected saddle in addition to the disconnected and the totally connected ones,
we expect that the capacity of negativity exhibits several peaks around each phase transition.
In section \ref{Sec:capacity} we will see that this is indeed the case.

Before closing this section,
we present the explicit forms of the refined R\'{e}nyi negativity and the capacity of negativity for later use:
\begin{align}
    S^{\mathrm{T_B} (n)} &= - \sum_i \frac{(\la_i)^n}{Z^{\mathrm{T_B}(n)}} \log \frac{|\la_i|^n}{Z^{\mathrm{T_B}(n)}}, \label{ptS} \\
    C^{\mathrm{T_B} (n)} &=
    \sum_i \frac{(\la_i)^n}{Z^{\mathrm{T_B}(n)}} \left( \log \frac{|\la_i|^n}{Z^{\mathrm{T_B}(n)}} \right)^2 - \left( S^{\mathrm{T_B} (n)} \right)^2. \label{ptC}
\end{align}
Note that these formulas are valid for both even and odd $n$.
By using the eigenvalue density $D(\la)$ of $\rho_{\mathrm{AB}}^{\mathrm{T_B}}$ (also known as the negativity spectrum),
the sum $\sum_i$ in \eqref{ptS} and \eqref{ptC} can be replaced by the integral $\int d\la D(\la)$.
In section \ref{Sec:JT} and \ref{Sec:capacity}, we will study the refined R\'{e}nyi negativity
in the canonical ensemble and the capacity of negativity in the microcanonical ensemble 
using the negativity spectrum, respectively.

\section{JT gravity with dynamical anti-FZZT branes}\label{Sec:JT}
In this section, we review the matrix model of JT gravity with dynamical anti-FZZT branes
and explain our formalism to compute the effect of the backreaction of branes.
As shown in \cite{Okuyama:2021bqg},
the eigenvalue density of matrix model is deformed if we take account of the backreaction of branes.

\subsection{Matrix integral and black hole microstates}
Let us consider the JT gravity with $K$ anti-FZZT branes \cite{Okuyama:2021bqg},
which is described by the double scaling limit of the matrix integral
\begin{equation}
\begin{aligned}
   Z &= \int dH e^{- \Tr V(H)} \det (\xi + H)^{-K} \\
    &= \int dH dQ dQ^\dagger e^{- \Tr V(H) - \Tr Q^\dagger (\xi + H) Q}.
\end{aligned} 
\label{eq:Z-mat}
\end{equation}
Here, $H$ and $Q$ are $N \times N$ hermitian and $N \times K$ complex matrices, respectively.
The anti-FZZT branes correspond to the inverse of the determinant operators. 
Here we take the parameter $\xi$ to be common to all $K$ branes.
In the double scaling limit, we take the large $N$ limit and zoom in on the edge of the eigenvalue spectrum at the same time 
(see e.g. \cite{Ginsparg:1993is} for a review).
The genus counting parameter $g_s$ of the double scaled model is a certain renormalized version of $1/N$.
In this paper we will further take the 't Hooft limit
\begin{equation}\label{'tHooft}
    K \rightarrow \infty,~ g_s \rightarrow 0 ~~ \text{with} ~~ t \equiv g_s K ~~ \text{fixed}
\end{equation}
and compute quantities in the planar approximation.
That is, we will ignore all higher-order corrections in $g_s$ and $K^{-1}$.

As argued in \cite{Penington:2019kki}, we can use the matrix model \eqref{eq:Z-mat} as a toy model for the black hole evaporation.
We identify the Hilbert space of black hole $\cH_{\text{BH}}$ and the Hilbert space of Hawking radiation $\cH_\text{R}$
with the degrees of freedom of  $H$ and $Q$, respectively, i.e.
\begin{equation}
\begin{aligned}
 \dim\cH_{\text{BH}}=N,\quad \dim \cH_\text{R}=K.
\end{aligned} 
\end{equation}
We also regard the 't Hooft parameter $t$ in \eqref{'tHooft} as an analogue of ``time''
since the number of Hawking quanta increases as the evaporation  proceeds.
To study the entanglement negativity we divide $K$ into two parts: 
$K = K_1 K_2$.
In other words, we consider the bipartite system $\cH_\text{R}=\cH_1\otimes\cH_2$ with
\begin{equation}
\begin{aligned}
 \dim \cH_1=K_1,\quad \dim \cH_2=K_2.
\end{aligned} 
\end{equation}
In the notation of the previous section, $\cH_1$  and $\cH_2$
correspond to $\cH_\mathrm{A}$ and $\cH_\mathrm{B}$, respectively.

We denote the components of $H, Q$ by $H_{ab}, Q_{a i_1 i_2}$,
where $a,b = 1, \ldots, N$ are ``color'' indices and $i_1, j_1 = 1, \ldots, K_1, ~ i_2, j_2 = 1, \ldots, K_2$ are ``flavor'' indices.
The color degrees of freedom describe the bulk gravity
while the flavor degrees of freedom correspond to the interior partners of the early Hawking radiation.
We write the matrix element $H_{ab}$ as
\begin{equation}
    H_{ab} = \bra a | H | b \ket
\end{equation}
where $\{| a \ket\}_{a=1}^N$ form the orthonormal basis of the
$N$ dimensional Hilbert space.
From the variable $Q_{a i_1 i_2}$, we define the (canonical) thermal pure quantum state \cite{Sugiura:2013pla, Goto:2021mbt},
\begin{equation}\label{state}
    | \psi_{i_1 i_2} \ket = \sum_a e^{-\hf \bt H} | a \ket Q_{a i_1 i_2}
    = \sum_{a,b} | b \ket \left( e^{-\hf \bt H} \right)_{ba} Q_{a i_1 i_2}.
\end{equation}
Here the inverse temperature $\bt$ is identified with the 
length of the asymptotic boundary
of the two dimensional spacetime. $| \psi_{i_1 i_2} \ket$ plays the role of the black hole microstates.

\subsection{Ensemble average}
In the matrix model of JT gravity,
the entanglement negativity is obtained from the ensemble average of the overlap 
$\bra \psi_{i_1 i_2} | \psi_{j_1 j_2} \ket$
and its partial transpose.
The ensemble average of $\cO$ is defined by
\begin{equation}
    \bra \overline{\cO} \ket = \int dH dQ dQ^\dagger e^{- \Tr V(H) - \Tr Q^\dagger (\xi + H) Q} \cO.
\end{equation}
Here the angle bracket $\bra\cO\ket$ represents the averaging over the color degrees of freedom
while the overline $\overline{\cO}$ represents averaging over the flavor degrees of freedom.
It is convenient to change the variable as
\begin{equation}
    Q = (\xi + H)^{-\hf} C,
\end{equation}
so that the new random variable $C$ obeys the Gaussian distribution
\begin{equation}
    \bra \overline{\cO} \ket = \int dH dC dC^\dagger \det (\xi + H)^{-K} e^{- \Tr V(H) - \Tr C^\dagger C} \cO.
\end{equation}
Thus, the average over $C$ can be computed by the Wick contraction.
Note that the determinant factor 
$\det (\xi + H)^{-K}$ naturally arises in this change of variable.

\begin{figure}[h]
    \centering
\subcaptionbox{ $W_{i_1 i_2, j_1 j_2}$ \label{sfig:amp-a}}{
\includegraphics[keepaspectratio,scale=0.6,width=0.4\linewidth]{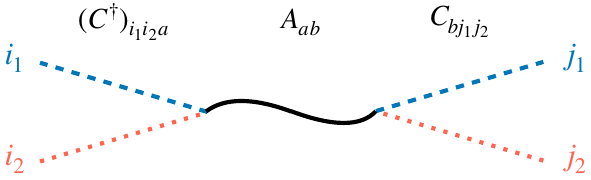}}
\hskip8mm
\subcaptionbox{$\left( W^\mathrm{T_2} \right)_{i_1 i_2, j_1 j_2}$\label{sfig:amp-b}}{\includegraphics[keepaspectratio,scale=0.6,width=0.4\linewidth]{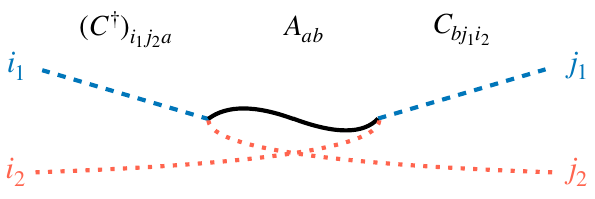}}
\caption{The overlaps (\ref{overlap}) and (\ref{trans-overlap}) are represented by the diagrams \subref{sfig:amp-a} and \subref{sfig:amp-b}, respectively.
Here, the black thick curve labeled by the color matrix $A_{ab}$ corresponds to the asymptotic boundary of
two dimensional spacetime while the dashed and dotted lines correspond to the first and second flavor degrees of freedom, respectively.}
\label{amplitude}
\end{figure}

On the other hand, the thermal pure state (\ref{state}) becomes
\begin{equation}
    | \psi_{i_1 i_2} \ket = \sum_{a,b} | b \ket \left[ e^{-\hf \bt H} (\xi + H)^{-\hf} \right]_{ba} C_{a i_1 i_2}
    = \sum_{a,b} | b \ket \left( \sqrt{A} \right)_{ba} C_{a i_1 i_2},
\end{equation}
where $A(H)$ is defined by
\begin{equation}
    A(H) = \frac{e^{-\bt H}}{\xi + H}.
\end{equation}
As explained in \cite{Penington:2019kki}, the reduced density matrix of the Hawking radiation is given by
\begin{equation}
    \rho_{i_1 i_2, j_1 j_2} = \frac{W_{i_1 i_2, j_1 j_2}}{\Tr W},
\end{equation}
where $W$ is given by
\begin{equation}\label{overlap}
    W_{i_1 i_2, j_1 j_2} \equiv \bra \psi_{i_1 i_2} | \psi_{j_1 j_2} \ket
    = \sum_{a,b} A_{ab} C_{a i_1 i_2}^\ast C_{b j_1 j_2}.
\end{equation}
To compute the entanglement entropy of the Hawking radiation, we need the average of $\Tr \rho^n$.
For instance, the flavor average of $W$ is obtained by using the Wick contraction
$\overline{C_{a i_1 i_2}^\ast C_{b j_1 j_2}} = \delta_{ab} \delta_{i_1 j_1} \delta_{i_2 j_2}$ as
\begin{equation}
    \overline{W_{i_1 i_2, j_1 j_2}} = \delta_{i_1 j_1} \delta_{i_2 j_2} \Tr A.
\end{equation}
As discussed in \cite{Penington:2019kki}, one can visualize the above computation by drawing diagrams.
For instance, the overlap (\ref{overlap}) is represented by the diagram in Figure
\ref{sfig:amp-a}.

To compute the entanglement negativity,
we take the partial transpose $\mathrm{T_2}$ of $W$ 
on the second flavor factor $\cH_2$
\begin{equation}\label{trans-overlap}
    \left( W^\mathrm{T_2} \right)_{i_1 i_2, j_1 j_2} =  W_{i_1 j_2, j_1 i_2}
    =  \sum_{a,b} A_{ab} C_{a i_1 j_2}^\ast C_{b j_1 i_2},
\end{equation}
and the partially transposed density matrix is written as
\begin{equation}
    \left( \rho^\mathrm{T_2} \right)_{i_1 i_2, j_1 j_2}
    = \frac{\left( W^\mathrm{T_2} \right)_{i_1 i_2, j_1 j_2}}{\Tr W}
    = \frac{1}{\Tr W} \sum_{a,b} A_{ab} C_{a i_1 j_2}^\ast C_{b j_1 i_2}.
\end{equation}
This partial transpose is represented by the diagram in Figure \ref{sfig:amp-b}.
In a similar manner as above,
we can compute the flavor average of $W^\mathrm{T_2}$
by using the Wick contraction
\begin{equation}
    \overline{\left( W^\mathrm{T_2} \right)_{i_1 i_2, j_1 j_2}} = \delta_{i_1 j_1} \delta_{j_2 i_2} \Tr A.
\end{equation}
Note that the flavor average of $W^\mathrm{T_2}$ is equal to that of $W$.

\begin{figure}[h]
    \centering
    \begin{minipage}[c]{0.2\hsize}
        \centering
        \includegraphics[keepaspectratio, scale=0.2]{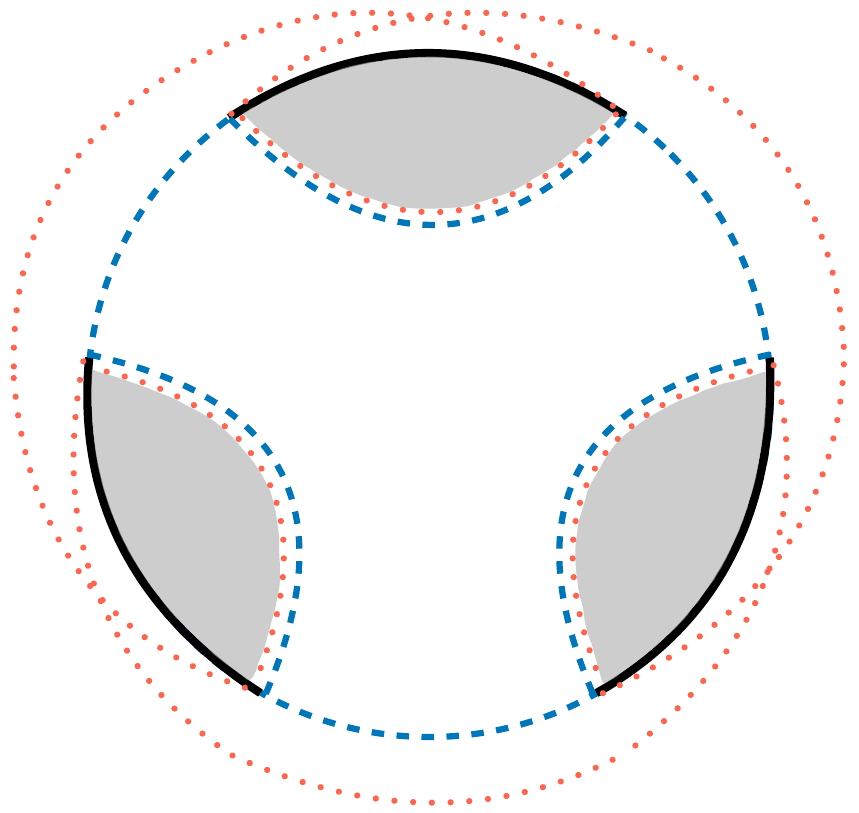}
        \subcaption*{$g = \mathbb{1}$}
    \end{minipage}
    \begin{minipage}[c]{0.2\hsize}
        \centering
        \includegraphics[keepaspectratio, scale=0.2]{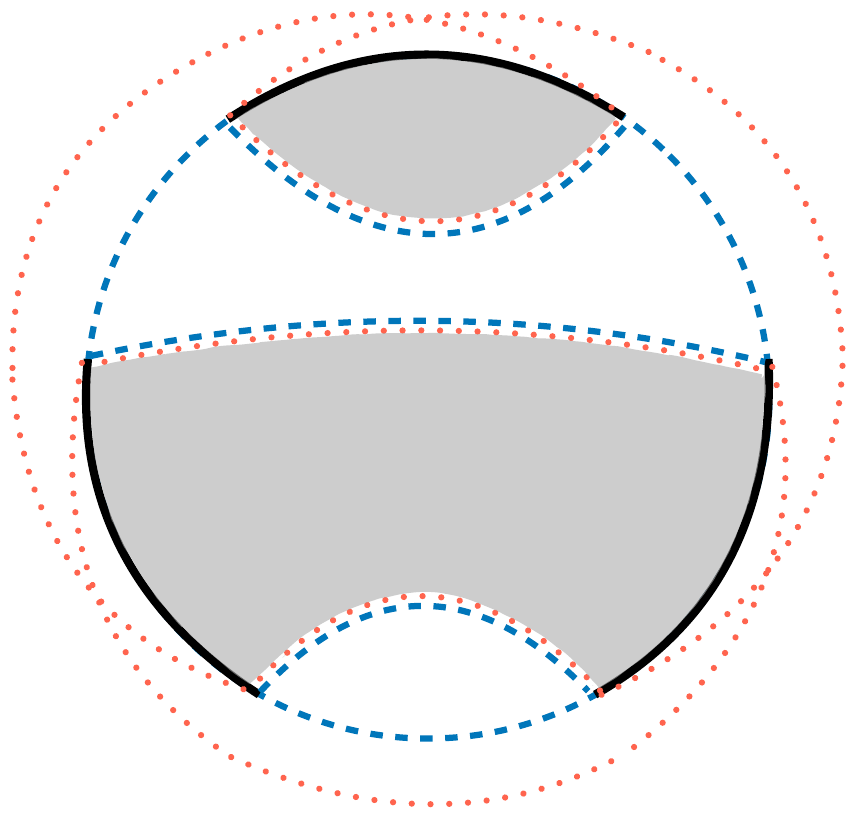}
        \subcaption*{$g = \tau$}
    \end{minipage}
    \begin{minipage}[c]{0.2\hsize}
        \centering
        \includegraphics[keepaspectratio, scale=0.2]{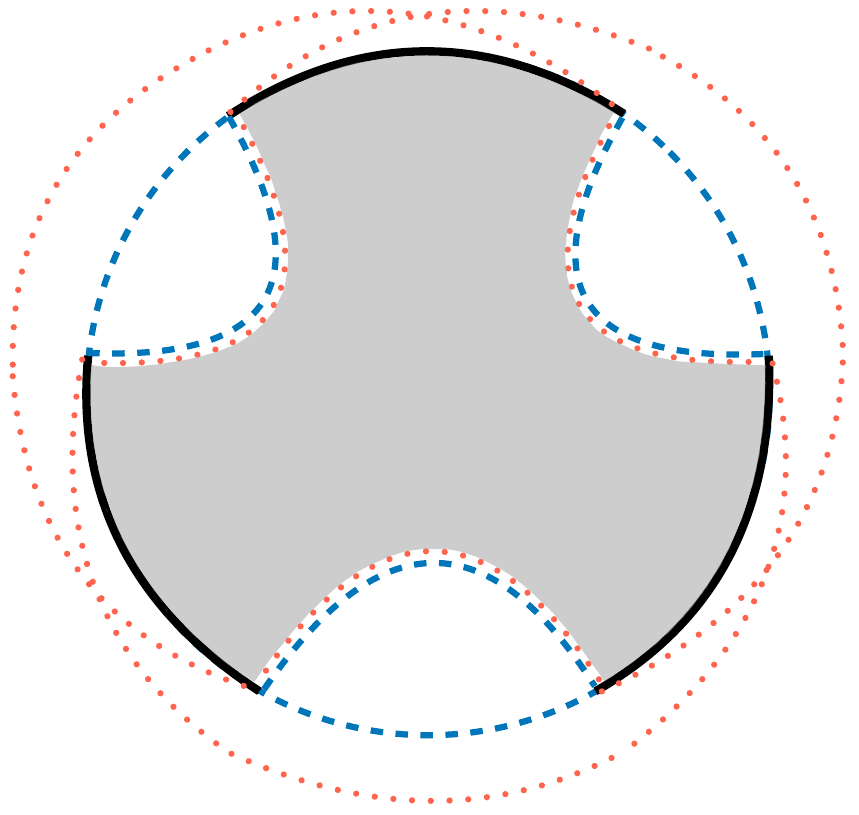}
        \subcaption*{$g = X$}
    \end{minipage}
    \begin{minipage}[c]{0.2\hsize}
        \centering
        \includegraphics[keepaspectratio, scale=0.2]{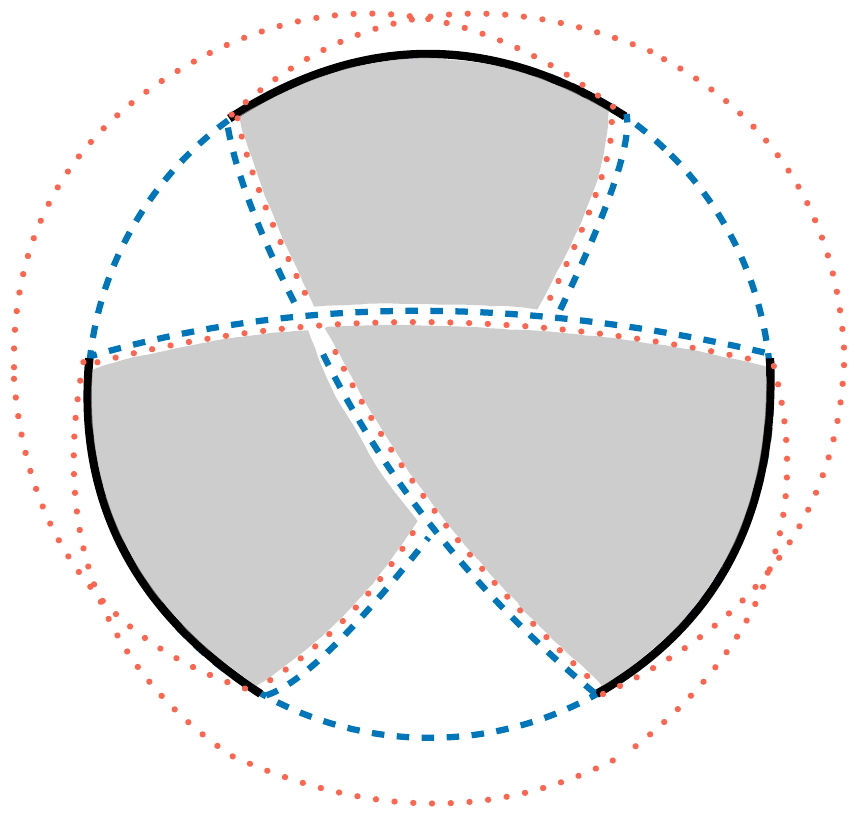}
        \subcaption*{$g = X^{-1}$}
    \end{minipage}
\caption{
The four dominant saddles of (\ref{permutation-sum}) are represented by these diagrams.
They are the examples of $n = 3$ case.
Taking the trace over first and second flavor degrees of freedom corresponds to the dashed and dotted loops respectively,
and the $\Tr A^m$ is represented by the gray colored region with $m$ asymptotic boundaries.}
\label{saddles}
\end{figure}

In the planar approximation, we can take the average of the numerator and the denominator independently
\begin{equation}
    \overline{\Tr \left( \rho^\mathrm{T_2} \right)^n} \approx
    \frac{\overline{\Tr \left( W^\mathrm{T_2} \right)^n}}{\left(\overline{\Tr W}\right)^n}.
\end{equation}
For instance, the results of $n = 2, 3$ read
\begin{align}
    \overline{\Tr \left( \rho^\mathrm{T_2} \right)^2} &\approx \frac{K (\Tr A)^2 + K^2 \Tr A^2}{(K \Tr A)^2}, \\
    \overline{\Tr \left( \rho^\mathrm{T_2} \right)^3} &\approx
    \frac{K (\Tr A)^3 + 3 K^2 \Tr A^2 \Tr A +  K K_2^2 \Tr A^3 + K K_1^2 \Tr A^3}{(K \Tr A)^3}.
\end{align}
For general $n$, the flavor average of $\Tr \left( \rho^\mathrm{T_2} \right)^n$
is expressed as a sum over the permutation group $S_n$ \cite{Dong:2021oad}
\begin{equation}\label{permutation-sum}
    \overline{\Tr \left( \rho^\mathrm{T_2} \right)^n} \approx
    \frac{1}{(K \Tr A)^n} \sum_{g \in S_n} \left( \prod_{i = 1}^{\chi (g)} \Tr A^{|c_i (g)|} \right)
    K_1^{\chi (g^{-1} X)} K_2^{ \chi (g^{-1} X^{-1})},
\end{equation}
where $\chi (g)$ is the number of disjoint cycles of the permutation $g$,
$|c_i (g)|$ is the length of the $i$-th disjoint cycle of $g$,
and $X ~ (X^{-1})$ is the (anti-)cyclic permutation of length $n$.
Each term of the sum \eqref{permutation-sum}
corresponds to a replica wormhole with the connection 
of $n$ replicas specified by $g\in S_n$.
%
As discussed in \cite{Dong:2021oad}, there are four dominant saddles 
in \eqref{permutation-sum} in the 't Hooft limit:
the disconnected saddle $g = \mathbb{1}$, the cyclically connected saddle $g = X$,
the anti-cyclically connected saddle $g = X^{-1}$, and the pairwise connected saddle $g = \tau$.
They are represented diagramatically in Figure \ref{saddles}.
Note that the pairwise connected saddle $g = \tau$ breaks the replica symmetry, while other saddles are replica symmetric.

Next, we consider the ensemble average over the color degrees of freedom.
Again, in the planar approximation we can take the average of the numerator and the denominator independently
\begin{equation}\label{n-moment}
    \bra \overline{\Tr \left( \rho^\mathrm{T_2} \right)^n} \ket \approx
    \frac{1}{(K \bra \Tr A \ket)^n} \sum_{g \in S_n} \left( \prod_{i = 1}^{\chi (g)} \bra \Tr A^{|c_i (g)|} \ket \right)
    K_1^{\chi (g^{-1} X)} K_2^{ \chi (g^{-1} X^{-1})}.
\end{equation}
Finally, the problem boils down to the computation of $Z_n$ defined by
\begin{equation}\label{path-int}
   Z_n\equiv \bra \Tr A^n \ket = \int dH \det (\xi + H)^{-K} e^{- \Tr V(H)} \Tr A^n.
\end{equation}
In this paper we consider the JT gravity matrix model, which corresponds to a particular background $t_k = \ga_k$ of topological gravity \cite{Okuyama:2019xbv} with
\begin{equation}
    \ga_0 = \ga_1 = 0, ~~ \ga_k = \frac{(-1)^k}{(k - 1)!} ~~ (k \geq 2).
\end{equation}
As discussed in \cite{Okuyama:2021eju}, in the 't Hooft limit \eqref{'tHooft}
the insertion of anti-FZZT branes strongly backreacts to the eigenvalue density, 
which leads to a shift of the couplings $t_k$ of topological gravity
\begin{equation}
    t_k = \ga_k + t (2k - 1)!! (2 \xi)^{-k - \hf}.
\end{equation}
In the planar approximation,
(\ref{path-int}) is written in terms of the genus-zero eigenvalue density $\rho_0 (E)$ of $H$ as
\begin{equation}\label{planar-pi}
    Z_n = \int_{E_0}^{\infty} dE \rho_0 (E) A(E)^n,
\end{equation}
where $\rho_0 (E)$ is given by
\begin{equation}\label{density}
    \rho_0 (E) = \frac{1}{\sqrt{2} \pi g_s} \left( \int_{E_0}^E dv \frac{I_0 (2 \sqrt{v})}{\sqrt{E - v}}
    - \frac{t}{E + \xi} \sqrt{\frac{E - E_0}{2 (E_0 + \xi)}} \right).
\end{equation}
Here the threshold energy $E_0$ is defined by the largest negative solution of the genus-zero string equation
\begin{equation}
    \sqrt{E_0} I_1 (2 \sqrt{E_0}) + \frac{t}{\sqrt{2 \xi + 2 E_0}} = 0,
\label{eq:string-eq}
\end{equation}
where $I_k (z)$ is the modified Bessel function of the first kind.
Note that there is a critical point $t = t_c$ beyond which 
the string equation \eqref{eq:string-eq} ceases to have a real solution of $E_0$.
As discussed in \cite{Okuyama:2021bqg}, the time scale $t=t_c$ can be interpreted as the end-point of the Hawking evaporation.

As a comparison, we comment on the result in the probe brane approximation.
The color average of $\Tr A^n$ in the probe brane approximation is given by
\begin{equation}\label{probe-pi}
    \bra \Tr A^n \ket_\text{probe} 
    = \int_0^\infty dE \rho_0^\text{JT} (E) A(E)^n,
\end{equation}
where the original JT gravity density of state $\rho_0^\text{JT} (E)$ is given by
\begin{equation}\label{probe-density}
    \rho_0^\text{JT} (E) = \frac{\mathrm{sinh}(2 \sqrt{E})}{\sqrt{2} \pi g_s}.
\end{equation}
We emphasize that we have treated anti-FZZT branes as dynamical objects.
More specifically, the color average in (\ref{path-int}) is evaluated in the presence of the determinant factor
$\det (\xi + H)^{-K}$ and as a consequence 
the eigenvalue density is deformed from $\rho_0^\text{JT} (E)$ in (\ref{probe-density}) to $\rho_0 (E)$ in (\ref{density}).
This is the main difference from the approach of \cite{Dong:2021oad}, which is based on the probe brane approximation.

In Table \ref{saddle-value}, we summarize the result of the R\'{e}nyi negativities (\ref{even-Renyi}) and (\ref{odd-Renyi})
obtained from \eqref{n-moment} for each dominant saddle $g=\mathbb{1},\tau,X,X^{-1}$ \cite{Dong:2021oad}.
To evaluate $Z_n$ in Table \ref{saddle-value}, we have to use the deformed eigenvalue density $\rho_0(E)$ as in \eqref{planar-pi}.
As we will see in section \ref{Sec:capacity}, the dominant saddles change as $\mathbb{1} \rightarrow \tau \rightarrow X$
when we increase $K_1$ with fixed $K_2$ and $g_s$.
\begin{table}[h]
    \renewcommand{\arraystretch}{1.2}
	\begin{center}
	\begin{tabular}{c|cccc}
    \hline
	$\displaystyle g$ &$\displaystyle \mathbb{1}$  &$\displaystyle \tau$ &$\displaystyle X$ &$\displaystyle X^{-1}$ \\ \hline \hline
{\footnotesize \begin{tabular}{c}
    $\displaystyle Z^{\mathrm{T_2}(2 m,\text{even})}$ \rule[-4mm]{0mm}{10mm} \\
    $\displaystyle Z^{\mathrm{T_2}(2 m - 1,\text{odd})}$ \rule[-4mm]{0mm}{10mm}
\end{tabular}}
&{\footnotesize \begin{tabular}{c}
    $\displaystyle \frac{1}{K^{2m-1}}$ \rule[-4mm]{0mm}{10mm} \\
    $\displaystyle \frac{1}{K^{2m-2}}$ \rule[-4mm]{0mm}{10mm}
\end{tabular}}
&{\footnotesize \begin{tabular}{c}
    $\displaystyle \frac{C_m Z_2^m}{K^{m-1} Z_1^{2m}}$ \rule[-4mm]{0mm}{10mm}\\
    $\displaystyle \frac{(2m-1) C_{m-1} Z_2^{m-1}}{K^{m-1} Z_1^{2m-2}}$ \rule[-4mm]{0mm}{10mm}
\end{tabular}}
&{\footnotesize \begin{tabular}{c}
    $\displaystyle \frac{Z_{2m}}{K_2^{2m-2} Z_1^{2m}}$ \rule[-4mm]{0mm}{10mm}\\
    $\displaystyle \frac{Z_{2m-1}}{K_2^{2m-2} Z_1^{2m-1}}$ \rule[-4mm]{0mm}{10mm}
\end{tabular}}
&{\footnotesize \begin{tabular}{c}
    $\displaystyle \frac{Z_{2m}}{K_1^{2m-2} Z_1^{2m}}$ \rule[-4mm]{0mm}{10mm}\\
    $\displaystyle \frac{Z_{2m-1}}{K_1^{2m-2} Z_1^{2m-1}}$ \rule[-4mm]{0mm}{10mm}
\end{tabular}} \\
    \hline
	\end{tabular}
	\end{center}
\caption{
The values of R\'{e}nyi negativities in each dominant saddle.
Here ${\footnotesize C_m = \frac{1}{m+1} \left( \begin{array}{c} 2m \\ m \end{array} \right) }$ is the Catalan number
which comes from the number of non-crossing pairings.
}
\label{saddle-value}
\end{table}

\subsection{Resolvent and negativity spectrum} \label{Sec:resolvent}
To compute $\bra \overline{\Tr \left( \rho^\mathrm{T_2} \right)^n} \ket $ in (\ref{n-moment}),
we need to evaluate the sum over the permutation group.
Following \cite{Dong:2021oad}, we use the resolvent method to compute the sum efficiently.
Here we define the resolvent $R(\la)$ by
\begin{equation}\label{resolvent}
    R(\la) = \Tr \left( \frac{1}{\la - \rho^\mathrm{T_2}} \right)
    = \sum_{n = 0}^\infty \frac{\Tr \left( \rho^\mathrm{T_2} \right)^n}{\la^{n+1}}
    = \frac{K}{\la} + \sum_{n = 1}^\infty \frac{\Tr \left( \rho^\mathrm{T_2} \right)^n}{\la^{n+1}}.
\end{equation}
The eigenvalue density $D(\la)$ of $\rho^\mathrm{T_2}$
can be obtained by taking the discontinuity of $R(\la)$ across the real axis
\begin{equation}\label{discontinuity}
    D(\la) = \lim_{\ep \rightarrow 0^+} \frac{R(\la - \ri \ep) - R(\la + \ri \ep)}{2 \pi \ri}
    = \lim_{\ep \rightarrow 0^+} \frac{1}{\pi} \mathrm{Im} R (\la - \ri \ep ).
\end{equation}
Note that $D(\la)$ is often called the negativity spectrum.
In what follows we will omit the sign of averaging $\bra\overline{\cO}\ket$ for notational simplicity.
Then, the average of R\'{e}nyi negativity (\ref{Renyi}) can be computed by
\begin{equation}\label{ave-moment}
    Z^{\mathrm{T_2}(n)} = \Tr \left( \rho^\mathrm{T_2} \right)^n 
    = \int_{-\infty}^\infty d\la D(\la) \la^n.
\end{equation}
Taking the derivative of (\ref{ave-moment}) with respect to
$n$,
the refined R\'{e}nyi negativity (\ref{ptS}) and the capacity of negativity (\ref{ptC}) are given by
\begin{align}
    S^{\mathrm{T_2} (n)} &= - \int_{-\infty}^\infty d\la D(\la)
    \frac{\la^n}{Z^{\mathrm{T_2}(n)}} \log \frac{|\la|^n}{Z^{\mathrm{T_2}(n)}}, \label{ave-entropy} \\
    C^{\mathrm{T_2} (n)} &= \int_{-\infty}^\infty d\la D(\la) \frac{\la^n}{Z^{\mathrm{T_2}(n)}}
    \left( \log \frac{|\la|^n}{Z^{\mathrm{T_2}(n)}} \right)^2 - \left( S^{\mathrm{T_2} (n)} \right)^2. \label{ave-capacity}
\end{align}
Note that we defined
$S^{\mathrm{T_2} (n)}$ and $C^{\mathrm{T_2} (n)}$ in terms of the
so-called annealed free energy $\log \bra \overline{Z^{\mathrm{T_2}(n)}} \ket$.
One can consider the quenched free energy $\bra \overline{\log Z^{\mathrm{T_2}(n)}}\ket$ as well,
but here we only consider the annealed quantities for simplicity.

As shown in \cite{Dong:2021oad},
the resolvent in (\ref{resolvent}) satisfies the following Schwinger-Dyson 
equation
\begin{equation}\label{Schwinger-Dyson}
    \la R(\la) = K + K_2^2 \int_{E_0}^{\infty} dE \rho_0(E)
    \frac{w(E) R(\la) \left( K + w(E) R(\la) \right)}{K^2 K_2^2 - w(E)^2 R(\la)^2},
\end{equation}
where $w(E) \equiv \frac{A(E)}{Z_1}$.
This equation is valid in the parameter regime $K_2 \ll K_1 g_s^{-1}$, and we will compute all quantities in this regime.
%
%
As discussed in \cite{Dong:2021oad}, we can solve the
Schwinger-Dyson equation (\ref{Schwinger-Dyson}) approximately by iteration.
First, we introduce $E_K$ and $\la_0$ by
\begin{align}
    K &= K_2^2 \int_{E_0}^{E_K} dE \rho_0(E),  \label{num_of_states} \\
    \la_0 &=\frac{1}{K} \int_{E_K}^{\infty} dE \rho_0(E) w(E), \label{lambda-zero}
\end{align}
and approximate the integral in (\ref{Schwinger-Dyson}) as
\begin{equation}\label{approx-resolvent}
\begin{aligned}
    \la R(\la) \approx K &+ K_2^2 \int_{E_0}^{E_K} dE \rho_0(E)
    \frac{w(E) R(\la) \left( K + w(E) R(\la) \right)}{K^2 K_2^2 - w(E)^2 R(\la)^2} \\
    &+ \frac{1}{K} \int_{E_K}^{\infty} dE \rho_0(E) w(E) R(\la).
\end{aligned}
\end{equation}
In the second line, we assumed the relation $K \gg |w(E) R(\la)|$ for $E\geq E_K$.
Using (\ref{lambda-zero}),
(\ref{approx-resolvent}) can be rewritten as
\begin{equation}
    R(\la) \approx \frac{K}{\la - \la_0} + \frac{K_2^2}{\la - \la_0}
    \int_{E_0}^{E_K} dE \rho_0(E)
    \frac{w(E) R(\la) \left( K + w(E) R(\la) \right)}{K^2 K_2^2 - w(E)^2 R(\la)^2}.
\label{eq:ite-1}
\end{equation}
We can solve $R(\la)$ by the iteration starting from
\begin{equation}
    R(\la) \approx \frac{K}{\la - \la_0}.
\label{eq:ite-0}
\end{equation}
Plugging \eqref{eq:ite-0} back into \eqref{eq:ite-1}, 
we get the second order iteration
\begin{equation}
    R(\la) \approx \frac{K}{\la - \la_0} + \frac{1}{\la - \la_0} \int_{E_0}^{E_K} dE \rho_0(E)
    \frac{w(E) \left( \la - \la_0 + w(E) \right)}{ (\la - \la_0)^2 - \left(\frac{w(E)}{K_2}\right)^2}.
\end{equation}
Using (\ref{num_of_states}), we arrive at the following expression of the resolvent
\begin{equation}
\begin{aligned}
    R(\la)
    &\approx \frac{1}{\la - \la_0} \left[ K_2^2 \int_{E_0}^{E_K} dE \rho_0(E) + \int_{E_0}^{E_K} dE \rho_0(E)
        \frac{w(E) \left( \la - \la_0 + w(E) \right)}{ (\la - \la_0)^2 - \left(\frac{w(E)}{K_2}\right)^2} \right] \\
    &= \int_{E_0}^{E_K} dE \rho_0(E) \left[
        \frac{K_2(K_2 + 1)}{2 \left( \la - \la_0 - \frac{w(E)}{K_2}\right)}
        + \frac{K_2(K_2 - 1)}{2 \left( \la - \la_0 + \frac{w(E)}{K_2} \right)} \right].
\end{aligned}
\end{equation}
Note that this solution is consistent with the large $\la$ behavior of the resolvent $R(\la) = K/\la+\cO(\la^{-2})$. 
Finally, taking the discontinuity across the real axis (\ref{discontinuity}), we find the negativity spectrum
\begin{equation}\label{canonical-spectrum}
    D(\la) = \int_{E_0}^{E_K} dE \rho_0(E) \left[
        \frac{K_2(K_2 + 1)}{2} \cob \left( \la - \la_0 - \frac{w(E)}{K_2}\right)
        + \frac{K_2(K_2 - 1)}{2} \cob \left( \la - \la_0 + \frac{w(E)}{K_2} \right) \right].
\end{equation}
We can check that the negativity spectrum in (\ref{canonical-spectrum}) satisfies the correct normalization conditions
\begin{equation}
    \int_{-\infty}^\infty d\la D(\la) = K, ~~~
    \int_{-\infty}^\infty d\la D(\la) \la = 1,
\end{equation}
which correspond to $\Tr 1=K$ and $\Tr \left( \rho^\mathrm{T_2} \right) = 1$, respectively.
In the next section, we will numerically study
the refined R\'{e}nyi negativity using the negativity spectrum in (\ref{canonical-spectrum}).

\section{Page curve for the refined R\'{e}nyi negativity}\label{Sec:Page}
In this section, we study the refined R\'{e}nyi negativity $S^{\mathrm{T_2}(n)}$
by including the effect of the backreaction of branes.
We consider $S^{\mathrm{T_2}(n)}$ as a function of $K_1$ with fixed $K_2$ and $g_s$.
We find numerically that $S^{\mathrm{T_2}(n)}$ decreases monotonically at late time of the evaporation.
It turns out that this decreasing behavior of $S^{\mathrm{T_2}(n)}$
can be proved analytically
in the large $\xi$ limit
by using the relation (\ref{thrmo-negativity}) with the refined R\'{e}nyi entropy.

\begin{figure}[h]
    \centering
    \includegraphics[keepaspectratio, scale=1.4]{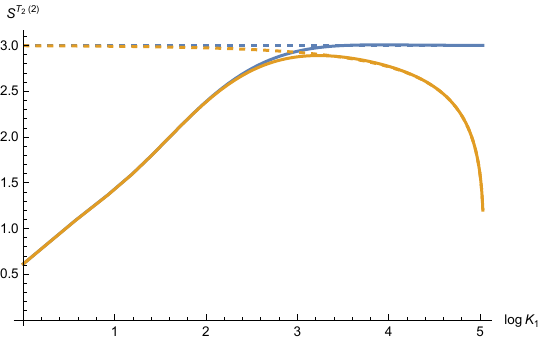}
\caption{
Plot of the refined R\'{e}nyi negativity (\ref{ave-entropy}) for $n = 2$ as a function of $\log K_1$.
The solid orange and blue curves are dynamical and probe brane cases, respectively.
We set $\xi = 50, \bt = 4, g_s = 1/50, K_2 = 2$ in this figure.
The ``thermodynamic negativity'' (\ref{thrmo-negativity}) are represented by the dashed lines.
}
\label{plot:negativity}
\end{figure}

Let us study the behavior of $S^{\mathrm{T_2}(n)}$ for $n=2$ as a typical example.
By using (\ref{ave-moment}) and (\ref{ave-entropy}) with (\ref{canonical-spectrum}),
we can compute the refined R\'{e}nyi negativity $S^{\mathrm{T_2}(n)}$ numerically.
Here we use the density of states (\ref{density}) including the effect of the backreaction of branes.
As a comparison, we also compute $S^{\mathrm{T_2}(n)}$ using the density of states (\ref{probe-density})
in the probe brane approximation.
As we can see from Figure \ref{plot:negativity}, in the dynamical brane case (orange curve)
 $S^{\mathrm{T_2}(n)}$ exhibits a monotonically decreasing behavior at late time of the evaporation,
while $S^{\mathrm{T_2}(n)}$ approaches a constant value in the probe brane approximation (blue curve).

As advertised above, we can prove the decreasing behavior of
$S^{\mathrm{T_2}(n)}$ analytically in the large $\xi$ limit
using the result of \cite{Okuyama:2021bqg}.
At late time, the totally connected saddle $g = X$ gives a dominant contribution to $S^{\mathrm{T_2}(n)}$.
Thus, at late time we can replace $S^{\mathrm{T_2}(n)}$ by its totally connected part,
which we call the ``thermodynamic negativity.''
As shown in Table \ref{saddle-value}, the totally connected part of $Z^{\mathrm{T_2}(n)}$ is given by
\begin{equation}
    Z^{\mathrm{T_2}(n)} |_{g=X} = \frac{Z_n}{K_2^{n-f(n)} (Z_1)^n}, ~~~
    f(n) \equiv
    \begin{cases}
        1, & n ~\text{odd}, \\
        2, & n ~\text{even}.
    \end{cases}    
\end{equation}
Thus by using (\ref{even-pS}) and (\ref{odd-pS}), the thermodynamic negativity is given by
\begin{equation}\label{thrmo-negativity}
    S_\mathrm{thermo}^{\mathrm{T_2}(n)} = S_\mathrm{thermo}(n) + f(n) \log K_2,
\end{equation}
where $S_\mathrm{thermo}(n)$ is the ``thermodynamic entropy'' defined
by \cite{Nakaguchi:2016zqi, Dong:2016fnf}
\begin{equation}\label{thermo-entropy}
    S_\text{thermo} (n) = - n^2 \del_n \left[ \frac{1}{n} \log \frac{Z_n}{(Z_1)^n} \right].
\end{equation}
The thermodynamic entropy (\ref{thermo-entropy}) is the totally connected part of the refined R\'{e}nyi entropy.
As proved analytically in \cite{Okuyama:2021bqg},
$S_\text{thermo}(n)$ decreases monotonically as a function of $t=g_sK$
in the large $\xi$ limit.
From \eqref{thrmo-negativity},
this immediately implies that $S^{\mathrm{T_2}(n)}_\text{thermo}$
decreases monotonically as a function of $K_1$ with fixed $K_2$ and $g_s$.
Thus we conclude that $S^{\mathrm{T_2}(n)}$ decreases at late time since
the dominant contribution $S^{\mathrm{T_2}(n)}_\text{thermo}$ does.
In Figure \ref{plot:negativity}, $S^{\mathrm{T_2}(n)}_\text{thermo}$
is plotted by the dashed orange curve, which clearly approaches  
$S^{\mathrm{T_2}(n)}$ (solid orange curve) at late time.

Finally, we comment on an interesting observation about (\ref{thrmo-negativity})
related to the bulk entanglement wedge cross section \cite{Takayanagi:2017knl, Nguyen:2017yqw}
which is defined by the minimal cross section of the entanglement wedge in the 
semi-classical picture.\footnote{
In two-dimensional gravities such as JT gravity, the area of 
the entanglement wedge cross section is not well-defined since
the constant time-slice of the bulk is one-dimensional
and the entanglement wedge cross section becomes a point.
In JT gravity, the area of a point is replaced by 
the value of dilaton at that point.}
As discussed in \cite{Tamaoka:2018ned}, a quantity defined by
\begin{equation}\label{cross-section}
    \cE_W = S^\mathrm{T_B} (\rho_\mathrm{AB}) - S_\text{vN} (\rho_\mathrm{AB})
\end{equation}
can be identified with the bulk entanglement wedge cross section $E_W$ in the semi-classical limit.
Here $S^\mathrm{T_B} (\rho_\mathrm{AB})$ and $S_\text{vN} (\rho_\mathrm{AB})$ denote
the $n \rightarrow 1$ limit of the refined R\'{e}nyi negativity and the von Neumann entropy, respectively.
It is known that
the entanglement wedge cross section $E_W$ and the entanglement entropies $S_\mathrm{A}, S_\mathrm{B}$
satisfy the following inequalities \cite{Takayanagi:2017knl}
\begin{equation}\label{inequality}
    E_W (\rho_\mathrm{A B}) \leq \min \left\{ S_\mathrm{A}, S_\mathrm{B} \right\}
    \leq \log \min \left\{ \dim \cH_\mathrm{A}, \dim \cH_\mathrm{B} \right\}.
\end{equation}
In our case we consider a bipartite radiation system $\cH_\mathrm{R} = \cH_1 \otimes \cH_2$.
At late time the totally connected part of $\cE_W$ is dominant and 
it is given by the $n \rightarrow 1$ limit of (\ref{thrmo-negativity})
\begin{equation}\label{latetime-EWCS}
    \cE_W |_{g=X} = \lim_{n \rightarrow 1} \left[ S_\mathrm{thermo}^{\mathrm{T_2}(n)} - S_\text{thermo} (n) \right] = \log K_2,
\end{equation}
which saturates the bound in (\ref{inequality}).
This means that the radiation system 
$\cH_\mathrm{R} = \cH_1 \otimes \cH_2$ is maximally entangled at late time.
We have obtained this result in the toy model of JT gravity with branes.
However, we speculate that this is a general feature of black holes in arbitrary dimensions: 
Hawking quanta become maximally entangled at late time of the evaporation.

As we can see from Figure \ref{plot:negativity}, 
the refined R\'{e}nyi negativity does not exhibit a clear signal of the 
phase transition between different saddles labeled by $g=\mathbb{1},\tau,X$.
It turns out that the capacity of negativity is a more better 
indicator of the phase transitions, as we will see in the next section.

\section{Capacity of negativity}\label{Sec:capacity}
In this section, we study the capacity of negativity in the microcanonical ensemble.
We find that the capacity of negativity has two peaks around the phase transitions
as a function of $K_1$ with fixed $K_2$.
\footnote{
The capacity of negativity does not exhibit any peaks in the approximation of (\ref{canonical-spectrum}).
Here we instead consider the microcanonical ensemble in which the Schwinger-Dyson equation for the resolvent can be solved analytically.
}
Moreover, we find that, away from the peaks, 
the capacity of negativity approaches a universal constant value which only depends on the replica number.

\subsection{Microcanonical ensemble}
To compute the capacity of negativity (\ref{ave-capacity}),
we first review the negativity spectrum $D(\la)$ and its phase transitions in the microcanonical ensemble \cite{Dong:2021oad}.
We focus on some small energy window $[ E, E + \Delta E ]$ and introduce the microcanonical variables as
\begin{equation}\label{micro-value}
    e^{\bm S} = \rho(E) \Delta E, ~~
    \bm{Z}_n = \rho(E) A(E)^n \Delta E, ~~
    \bm{w}(E) = \frac{A(E)}{\bm{Z}_1} = e^{- \bm S}.
\end{equation}
Then the Schwinger-Dyson equation \eqref{Schwinger-Dyson} reduces to a cubic equation
\begin{equation}\label{micro-Schwinger-Dyson}
    R(\la)^3 + \left( \frac{e^{\bm S} K_2^2 - K}{\la} \right) R(\la)^2
    + e^{2 \bm{S}} K K_2^2 \left( \frac{1}{\la} - K \right) R(\la)
    + \frac{e^{2 \bm{S}} K^3 K_2^2 }{\la} = 0.
\end{equation}
This equation can be simplified as
\begin{equation}
    z G(z)^3 + (\bt - 1) G(z)^2 + (\al - z) G(z) + 1 = 0,
\label{cube-SD}
\end{equation}
where we defined the rescaled variables as
\begin{equation}
    z = K_2 e^{\bm S} \la, ~~
    G(z) = \frac{e^{- \bm S} R(\la)}{K K_2},~~
    \al = \frac{e^{\bm S}}{K_1}, ~~ 
\bt = \frac{e^{\bm S} K_2}{K_1}.
\end{equation}
This cubic equation \eqref{cube-SD} can be solved analytically \cite{Shapourian:2020mkc}
\begin{equation}\label{micro-solution}
    G(z) = \frac{e^{- \ri \theta} Q_1 (z)}{ \left( Q_2 (z) + \sqrt{P(z)} \right)^{1/3} }
            - e^{ \ri \theta} \left( Q_2 (z) + \sqrt{P(z)} \right)^{1/3} + \frac{1 - \bt}{3 z},
\end{equation}
where $\th = \pi / 3$ and,
\begin{align}
    Q_1 (z) &= \frac{3 z (\al - z) - (\bt - 1)^2}{9 z^2}, \\
    Q_2 (z) &= \frac{9 z (\bt - 1)(\al - z) - 27 z^2 - 2 (\bt - 1)^3}{54 z^3}, \\
    P (z) &= Q_1 (z)^3 + Q_2 (z)^2.
\end{align}

\begin{figure}[t]
    \centering
     \subcaptionbox{   $K_1 = e^5$ \label{sfig:Da}}{
  \includegraphics[keepaspectratio, scale=0.5,width=0.3\linewidth]{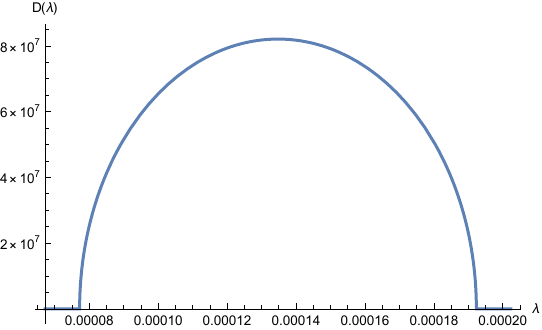} }
    \subcaptionbox{$K_1 = e^{10}$ \label{sfig:Db}}{   
\includegraphics[keepaspectratio, scale=0.5,width=0.3\linewidth]{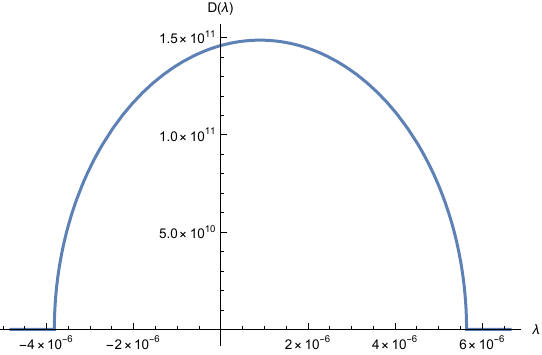} }
   \subcaptionbox{ $K_1 = e^{18}$  \label{sfig:Dc}}{   
\includegraphics[keepaspectratio, scale=0.5,width=0.3\linewidth]{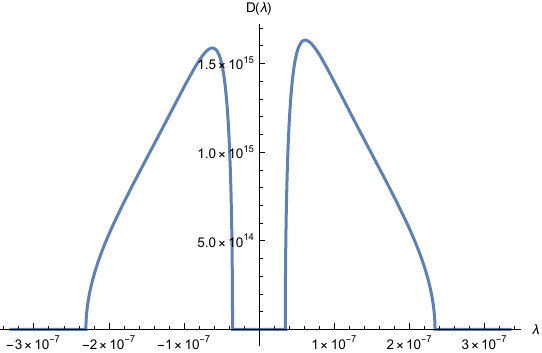}}
\caption{
Plot of the negativity spectrum obtained from
the discontinuity of $G(z)$ in (\ref{micro-solution}).
Here we set the parameters $\bm{S} = 12, K_2 = 50$.
\subref{sfig:Da}, \subref{sfig:Db}, \subref{sfig:Dc}  correspond to the dominant saddles $g = \mathbb{1}, \tau, X$, respectively.
}
\label{micro-spectrum}
\end{figure}

From the solution of $G(z)$ in \eqref{micro-solution},
we can compute the negativity spectrum $D(\la)$
by taking the discontinuity of $G(z)$ as in (\ref{discontinuity}).
In what follows, we will consider $D(\la)$ as a function of $K_1$ with fixed $K_2$ and $\bm{S}$.
As we vary $K_1$, there appear three phases for $D(\la)$ corresponding to three saddles labeled by 
$g = \mathbb{1}, \tau, X$ \cite{Dong:2021oad}.
Let us take a closer look at the behavior of $D(\la)$ in each phase.
As shown in Figure \ref{sfig:Da}, when the saddle labeled by $g = \mathbb{1}$ is dominant
$D(\la)$ has a support only on the positive $\la$ axis.
In the phase where the $g=\tau$ saddle is dominant 
(see Figure \ref{sfig:Db}), $D(\la)$ has 
a support on the negative $\la$ region as well,
which means that the partially transposed 
density matrix $\rho^{\mathrm{T_2}}$ has negative eigenvalues
and the two systems $\cH_1$ and $\cH_2$  become entangled.
If we further increase $K_1$, we land on the phase where the $g=X$ saddle is dominant (see Figure \ref{sfig:Dc}).
In this case, the support of $D(\la)$ is a disjoint union of a negative 
$\la$ region and a positive $\la$ region.

Now let us consider the conditions to determine the support of $D(\la)$ 
and its phase transition points.
$D(\la)$ behaves as a fractional power in $\la$ near the edges of the support
and hence its derivative diverges there.
Thus we can determine the edges of the support of $D(\la)$ from 
the singularities of $dR(\la)/d\la$, since
$D(\la)$ is obtained from the discontinuity of $R(\la)$.
From (\ref{micro-solution}) we find that 
the support of $D(\la)$ is determined by the condition $P(z) \geq 0$, and
the zeros of $P(z)$ correspond to
the edges of $D(\la)$.
To find the transition points, it is convenient to
define the polynomial part of $P(z)$ by $f(z) = z^4 P(z)$.
The transition between $g=\mathbb{1}$ and $g=\tau$
occurs when the left edge of the support of $D(\la)$ approaches $\la=0$.
This condition is given by $f(0) = 0$, from which $K_1$ is determined as
\begin{equation}\label{1st-transition}
    K_1 = \frac{e^{\bm S} \left( K_2 - \sqrt{K_2^2 - 1} \right)}{2}.
\end{equation}
The transition between $g=\tau$ and $g=X$
corresponds to a point where the support of $D(\la)$ starts to split into
two disjoint regions.
This is determined by the condition that $f(z)$
has a double root.
In other words, the discriminant of $f(z)$ becomes zero 
at this transition point.
From the discriminant of $f(z)$,
we find that the critical value of 
$K_1$ is determined by the following cubic equation
\begin{equation}\label{2nd-transition}
    64 K_1^3 - 48 e^{\bm S} K_2 K_1^2 - \left( 15 K_2^2 - 27 \right) e^{2 \bm{S}} K_1
    - e^{3 \bm{S}} K_2^2 = 0.
\end{equation}
One can check that the large $K_2$ limit of 
these conditions (\ref{1st-transition})
and (\ref{2nd-transition}) are consistent
with the result of \cite{Dong:2021oad}.
In the next subsection we show 
that the capacity of negativity has two peaks around these transition points.

\subsection{Capacity of negativity}
Let us consider the capacity of negativity 
$C^{\mathrm{T_2}(n)}$ 
in the microcanonical ensemble
using the result of $D(\la)$ in the previous subsection.
From the definition of $C^{\mathrm{T_2}(n)}$ 
in the third column of Table \ref{analogy}, one can show that 
$C^{\mathrm{T_2}(n)}$ is written as
\begin{equation}\label{2nd-derivative-form}
    C^{\mathrm{T_2} (n)} = n^2 \del_n^2 \log Z^{\mathrm{T_2}(n)}.
\end{equation}
This is valid for both even and odd $n$.

In the saddle-point approximation, 
we can replace $Z^{\mathrm{T_2}(n)}$ by its saddle-point value  
summarized in Table \ref{saddle-value}.
In general, the saddle-point values  of $Z^{\mathrm{T_2}(n)}$ in 
Table \ref{saddle-value} take the form
\begin{equation}
\begin{aligned}
 Z^{\mathrm{T_2}(n)}=a(n)b^{n+c},
\end{aligned} 
\label{eq:Z-form}
\end{equation}
where $b$ and $c$ are $n$-independent constants. Plugging 
\eqref{eq:Z-form} into \eqref{2nd-derivative-form} we find
\begin{equation}
\begin{aligned}
 C^{\mathrm{T_2} (n)} = n^2 \del_n^2 \log a(n).
\end{aligned} 
\end{equation}
Namely, $b$ and $c$ in \eqref{eq:Z-form} do not contribute to
$C^{\mathrm{T_2} (n)}$.
For instance, $Z^{\mathrm{T_2}(n)}=1/K^{n-1}$ for the $g = \mathbb{1}$ saddle
(see the first column in Table \ref{saddle-value}),
which implies that
$C^{\mathrm{T_2} (n)}$ vanishes 
when the $g=\mathbb{1}$ saddle is dominant.
From the second column of Table \ref{saddle-value},
the saddle-point value of $C^{\mathrm{T_2} (n)}$ 
in the pairwise connected phase $g = \tau$  is given by
\begin{align}
    C^{\mathrm{T_2} (n,\text{even})} &= \frac{n^2}{4} \left[ \psi^{(1)} \left( \frac{n+1}{2} \right)
    - \psi^{(1)} \left( \frac{n}{2} + 2 \right) \right], \label{even-tau-value} \\
    C^{\mathrm{T_2} (n,\text{odd})} &= -1 + \frac{n^2}{4} \left[ \psi^{(1)} \left( \frac{n}{2} \right)
    - \psi^{(1)} \left( \frac{n+3}{2}\right) \right], \label{odd-tau-value}
\end{align}
where $\psi^{(m)} (z)$ is the polygamma function of order $m$,
which comes from the derivative of the Catalan number $C_{m}$ or $C_{m-1}$.
The first term $-1$ in \eqref{odd-tau-value} comes 
from the derivatives of the factor $(2m-1)$ in $Z^{\mathrm{T_2} (2m-1,\text{odd})} $.
Interestingly, 
the values of $C^{\mathrm{T_2} (n)}$  
in (\ref{even-tau-value}) and (\ref{odd-tau-value}) 
depend only on the replica number $n$ and they are independent
of the other parameters $K_1,K_2$ and $\bm{S}$.
This property comes from the 
fact that the parameter dependence is contained in the factor $b^{n+c}$
in \eqref{eq:Z-form} which does not contribute to 
$C^{\mathrm{T_2} (n)}$ as we saw above.

In the totally connected phase $g = X$, the saddle-point value of
$ C^{\mathrm{T_2} (n)}$ is given by
\begin{equation}
    C^{\mathrm{T_2} (n)} = n^2 \del_n^2 \log Z_n.
\label{eq:saddle-X}
\end{equation}
This is valid not only in the microcanonical ensemble but also in the canonical ensemble.
Note that $ Z_n$
in \eqref{eq:saddle-X} is given by $Z_n=\bra\Tr A^n\ket$, which 
depends on how we take the ensemble average $\bra\cdots\ket$.
In the microcanonical ensemble,
$C^{\mathrm{T_2} (n)}$
in \eqref{eq:saddle-X} vanishes since $Z_n$ in 
\eqref{micro-value} takes the form of $b^{n+c}$.

\begin{figure}[t]
    \centering
    \includegraphics[keepaspectratio, scale=1.4]{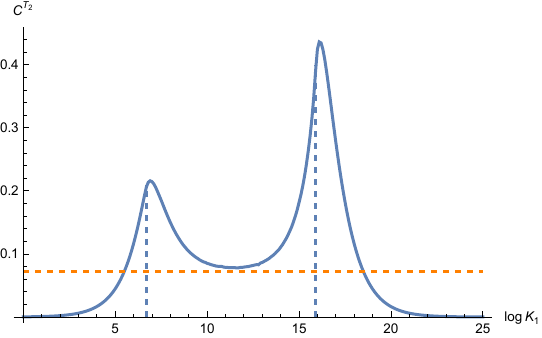}
\caption{
Plot of the capacity of negativity (\ref{ave-capacity}) for $n = 1$ as a function of $\log K_1$.
We set $K_2 = 50, \bm{S} = 12$ in this figure.
The two vertical dashed blue lines are the transition points determined by (\ref{1st-transition}) and (\ref{2nd-transition}).
The horizontal dashed orange line is the value of (\ref{odd-tau-value}) at $n = 1$.
}
\label{plot:capacity}
\end{figure}

Now let us consider the capacity of negativity $C^{\mathrm{T_2} (n)}$
in the microcanonical ensemble using the definitions in 
\eqref{ave-moment}, \eqref{ave-entropy} and \eqref{ave-capacity}.
In Figure \ref{plot:capacity} we show the plot of
$C^{\mathrm{T_2} (n)}$ for $n=1$ as a function of $K_1$ with fixed 
$K_2$ and $\bm{S}$.
As we see from Figure \ref{plot:capacity}, the capacity of negativity (blue curve) has two peaks around 
the transition points determined by the conditions (\ref{1st-transition}) and (\ref{2nd-transition})
(dashed vertical lines). 
Between the two peaks,
$C^{\mathrm{T_2} (n)}$ approaches a saddle-point value for $g = \tau$ (dashed orange line) given by (\ref{odd-tau-value}).
The locations of peaks do not exactly match the transition points 
determined by the conditions \eqref{1st-transition} and \eqref{2nd-transition}
since the capacity of negativity is an integrated function of $D(\la)$.
However, we have checked that
\eqref{1st-transition} and \eqref{2nd-transition} give a qualitatively
good approximation of the locations of peaks for various values 
of $K_2$, $\bm{S}$ and $n$.
From Figure \ref{plot:capacity}, we also observe that 
$C^{\mathrm{T_2} (n)}$ approaches zero in the small and large $K_1$ limits.
This is consistent with the fact that
the saddle-point values of $C^{\mathrm{T_2} (n)}$ vanish for 
$g=\mathbb{1}$ and $g=X$.

Near the transition points, the saddle-point approximation breaks down
and we cannot ignore the sub-leading corrections
in \eqref{permutation-sum}
coming from the general elements $g$ in the permutation group $S_n$
other than the dominant saddles $g=\mathbb{1},\tau,X$.
This is the physical origin of the peaks
of $C^{\mathrm{T_2} (n)}$ we observed in Figure \ref{plot:capacity}.
This is in contrast to the behavior of the refined R\'{e}nyi negativity
$S^{\mathrm{T_2} (n)}$
(see Figure \ref{plot:negativity})
which does not exhibit a clear signal of the phase transitions.
In $C^{\mathrm{T_2} (n)}$ the phase transitions manifest itself as
two peaks, which suggests that $C^{\mathrm{T_2} (n)}$
is a more better indicator of the phase transitions than $S^{\mathrm{T_2} (n)}$.
Indeed, in many physical systems
it is useful to consider the
susceptibility defined by a second derivative of the free energy
with respect to an 
external field in order to search for possible phase transitions. 
In our case, the capacity of negativity 
$C^{\mathrm{T_2} (n)}$ plays the role of the
susceptibility of entanglement in the bipartite system $\cH_1\otimes\cH_2$. 
In fact, $C^{\mathrm{T_2} (n)}$ in \eqref{2nd-derivative-form}
represents the variance of the partially transposed
modular Hamiltonian $H^{\mathrm{T_2}}=-\log \rho^{\mathrm{T_2}}$
\begin{equation}\label{fluctuation}
\begin{aligned}
 \frac{1}{n^2}C^{\mathrm{T_2} (n)}
 &=\bigl\bra(H^{\mathrm{T_2}})^2 \bigr\ket_n -\bigl\bra H^{\mathrm{T_2}}\bigr\ket_n^2 \\
 &=\bigl\bra(H^{\mathrm{T_2}} - \bigl\bra H^{\mathrm{T_2}}\bigr\ket_n)^2 \bigr\ket_n,
\end{aligned}
\end{equation}
where $\bra \cO\ket_n$ is defined by
\begin{equation}
\begin{aligned}
 \bra \cO\ket_n=
\frac{\Tr(\cO e^{-nH^{\mathrm{T_2}}})}{\Tr (e^{-nH^{\mathrm{T_2}}})}.
\end{aligned} 
\end{equation}
Near the phase transition the fluctuation of $H^{\mathrm{T_2}}$ becomes large,
which is observed as the peak of $C^{\mathrm{T_2} (n)}$.

\section{Conclusions and outlook}\label{Sec:conclusion}
In this paper,
we studied the refined R\'{e}nyi negativity and the capacity of negativity in the matrix model of JT gravity,
which serves as a toy model of the evaporating black hole.
First, we considered the refined R\'{e}nyi negativity in the presence of dynamical anti-FZZT branes.
The deformation of the density of states due to the backreaction of branes played an essential role in the computation.
As we see from Figure \ref{plot:negativity},
we find that the refined R\'{e}nyi negativity monotonically decreases at late time of the evaporation
by including the effect of the backreaction of branes, while it approaches a constant value in the probe brane approximation.
We can easily understand the decreasing behavior by using the 
relation \eqref{thrmo-negativity} between the totally connected part
of the refined R\'{e}nyi negativity 
and the refined R\'{e}nyi entropy.
The $n \rightarrow 1$ limit of the refined R\'{e}nyi negativity is related to the 
bulk entanglement wedge cross section in the semi-classical picture,
which saturates the inequality \eqref{inequality} 
at late time of the evaporation.
This means that the Hawking quanta become maximally entangled at late time of the evaporation.

Next, we considered the capacity of negativity in the microcanonical ensemble.
As we see from Figure \ref{plot:capacity}, we find that the capacity of negativity has two peaks around the phase transitions,
reflecting the fact that the entanglement negativity has more elaborate phase structure than that of the entanglement entropy.
This feature is the main difference from the capacity of entanglement.
Moreover, 
in the pairwise connected phase which arises in-between the two peaks,
the capacity of negativity approaches a universal constant value which only depends on the replica number.
We found that the capacity of negativity is an invaluable indicator of the phase transitions among the different entanglement phases.

There are many interesting open questions.
We studied the refined R\'{e}nyi negativity and the capacity of negativity in the planar approximation.
However, near the end of the evaporation the black hole becomes 
very small and we cannot ignore the quantum corrections to various quantities.
It would be interesting to study the higher genus corrections  to the entanglement negativities in JT gravity.
To this end 
we need to develop a method of computing the non-planar 
corrections to the resolvent of $\rho^{\mathrm{T_2}}$.
Moreover, it would be very interesting to compute the entanglement negativities non-perturbatively in $g_s$.

It would be interesting to study the capacity of negativity in the canonical ensemble.
We expect that the qualitative feature will also hold in the canonical ensemble:
the capacity of negativity exhibits two peaks around the phase transitions.
The main difference between the microcanonical and canonical ensembles is the late time behavior (\ref{eq:saddle-X}).
One of the advantages of the canonical ensemble is that we can study the quantities
using the density of states in JT gravity with the effect of the backreaction of branes.

Actually we tried to compute the capacity of negativity in the canonical ensemble.
However, it does not exhibit any peaks around the phase transitions by using the negativity spectrum (\ref{canonical-spectrum}).
On the other hand, there is a trick to find a peak of the capacity of entanglement in the canonical ensemble \cite{Kawabata:2021hac}.
Namely, we need to determine the minimal eigenvalue $\la_0$ of the entanglement spectrum
by solving the derivative of the Schwinger-Dyson equation $d\la/dR = 0$ numerically.
We studied the capacity of negativity in a similar manner.
However, it was difficult to find a solution of $d\la/dR = 0$
because the Schwinger-Dyson equation for the negativity resolvent is more complicated than that of the entanglement entropy.
We leave the computation of the capacity of negativity in the canonical ensemble as an interesting future problem.

We introduced the capacity of negativity as a natural analogue of the capacity of entanglement.
The authors of \cite{Kawabata:2021vyo} studied the capacity of entanglement
in a two dimensional dilaton gravity coupled to conformal matter with a large central charge.
This is a useful model where the island formula \cite{Penington:2019npb, Almheiri:2019psf, Almheiri:2019hni}
for the entanglement entropy of the Hawking radiation was 
explicitly tested for the first time.
Also, the island contributions to the entanglement negativity was studied in \cite{KumarBasak:2020ams}.
It would be interesting to consider the capacity of negativity in such a model
and to study the contributions of the island along the line of these studies.

It is also interesting to consider the holographic dual of the capacity of negativity.
As discussed in \cite{Nakaguchi:2016zqi, DeBoer:2018kvc},
the holographic dual of the capacity of entanglement is described by the graviton fluctuation around the minimal 
surface associated with holographic refined R\'{e}nyi entropy.
As we mentioned in section \ref{Sec:Page}, the refined R\'{e}nyi negativity is related to the bulk entanglement wedge cross section
which is the minimal cross section of the entanglement wedge in the 
semi-classical picture.
We speculate that the holographic dual of the capacity of negativity is described
by the graviton fluctuation around the entanglement wedge cross section.
Furthermore, the entanglement wedge cross section is also related to the reflected entropy \cite{Akers:2022max}.
It would be interesting to consider ``refined R\'{e}nyi reflected entropy'' and its capacity.
We expect that we can extract more detailed information about the entanglement structure
by using these 
generalization of quantum information quantities.

\acknowledgments
We would like to thank Tatsuma Nishioka for correspondence.
The work of KO was supported in part by MEXT-JSPS Grant-in-Aid for Transformative Research Areas (A) 
``Extreme Universe'' 21H05187 and JSPS KAKENHI Grant 22K03594.
TT is supported by JST SPRING, Grant Number JPMJSP2144 (Shinshu University).
\bibliography{references}
\bibliographystyle{utphys}

\end{document}